\documentclass[a4paper,12pt]{article}
\usepackage[utf8]{inputenc}
\usepackage{cite}
\usepackage{amsmath}
\usepackage{amssymb}
\usepackage{hyperref}
\usepackage{graphicx}
\usepackage{epstopdf}
\usepackage{makecell}

\date{}
\title{On the Dependence of Charge Density on Surface Curvature of an Isolated Conductor}
\author{Kolahal Bhattacharya\\kolahalb@tifr.res.in\\Tata Institute of Fundamental Research}

\begin{document}

\maketitle

\begin{abstract}
A study of the relation between the electrostatic charge density at a point on a conducting 
surface and the curvature of the surface (at that point) is presented. Two major papers in the scientific 
literature on this topic are reviewed and the apparent discrepancy between them is resolved. 
Hence, a step is taken towards obtaining a general analytic formula for relating the charge 
density with surface curvature of conductors. The merit of this formula and its limitations 
are discussed.
\end{abstract}

\section{Introduction}

It is an observational fact that on a conducting surface, electric charge density is greater at 
those points where local surface curvature is also higher. For example, if some amount of charge 
is given to a needle, the charge density would be very high at the sharp tip of the needle. The 
surface curvature is also very high there, compared to the smoother cylindrical portions of the 
needle. Although this is empirically known for a long time, the analytic formula connecting the 
two quantities is not known. A major development in this field may have a huge impact on modern 
experimental research areas, like scanning tunnelling microscopy, field emission cathode, field 
ion microscopy etc.

In general, the charge density at a point on a given conductor surface depends on the curvature 
of the surface at that point, the shape of the conductor and also on the presence or absence of 
the other nearby charged bodies. It is proportional to the magnitude of the electrostatic field 
just outside the conductor~\cite{griffiths1999introduction}. The field can be obtained from the 
electrostatic potential that is a solution to Laplace's equation, subjected to the boundary 
conditions~\cite{greiner2012classical}. However, the second order partial differential equation 
cannot be solved in general for an arbitrary configuration of charged bodies. Even for isolated 
charged conducting bodies, closed-form solutions are possible usually when the problems exhibit 
planar/cylindrical or spherical symmetries. However, if it is possible to solve, the charge 
density at any point on the conductor can be easily calculated. 

But this does not explicitly give the dependence of the charge density on the surface curvature. 
To obtain the desired dependence, one can start from the relation between the conductor surface 
curvature and the electric field near the surface. The corresponding equation was first derived 
by Green~\cite{green1889essay}, who showed that

\begin{equation}\label{1}
\frac{dE}{dn} = -2\kappa{E}= -E\left(\frac{1}{R_1}+\frac{1}{R_2}\right)
\end{equation}
where $\frac{dE} {dn}$ is the normal derivative of the electric field across the surface of the 
conductor and $\kappa=\frac{1}{2}(\frac{1}{R_1}+\frac{1}{R_2})$ is the local mean curvature ($R
_1$ and $R_2$ are the principal radii of curvature of the surface) at a given location. Sir J J 
Thomson gave recognition to this relation in 1891~\cite{bakhoum2008proof}. Various methods were 
used by different authors~\cite{estevez1985power},~\cite{pappas1986differential}, over the past 
century, to prove Eq.\eqref{1}. But its application to find a connection between charge density 
and conductor surface curvature was first performed by Luo Enze~\cite{enze1986distribution}. He 
found the expression for charge density as a function of the mean curvature of the surface. But 
that was not the end of the story. Liu~\cite{liu1987relation} (1987) showed that for conducting 
bodies of specific shapes, the charge density at a point on the surface was proportional to $0.2
5^{th}$ power to the Gaussian curvature of the surface at that point. McAllister verified Liu's 
observation~\cite{mcallister1990conductor} and argued that the result was correct only when the 
potential was a function of a single variable. Unaware of this, in his 1987 paper, Liu made the 
statement that `Results from conductors with surfaces of different shape are so consistent that 
it is natural for us to expect that the quantitative relation is a universal rule for conductors 
whose surfaces can be expressed by analytic functions'. Soon, this conjecture was proved to be 
incorrect by many authors~\cite{dube1989comment},~\cite{torres1989comment}.

Whereas the works by Luo and Liu enriched this field very much, there are some issues that need 
further discussions. Their works on the same problem led to apparently different results. Luo's 
expression for charge density as a function of the mean curvature bears no resemblance to Liu's 
result, which is in terms of the Gaussian curvature. Zhang criticised Luo's work by pointing to 
an incorrect assumption used for solving an integral~\cite{zhang1988comment}. On the other hand, 
Liu's formula cannot be true for surface with negative Gaussian curvature, as that will lead to 
imaginary value of the charge density. This is also not true for surfaces that have points with 
zero Gaussian curvatures. If it were so, then electric charge could not accumulate on any plane 
conducting surface. Therefore, there is enough scope of discussion of the results, known so far, 
for better understanding of the topic. 

In this article, the main results of the existing literature will be reviewed first in the next 
section~\ref{LuoLiu}. Then, in section~\ref{Bridge}, the connection between their works will be 
studied. In section~\ref{universal}, attempts will be made to obtain a general analytic formula 
to connect the charge density and the surface curvature of conductors. Hence, this formula will 
be used in section~\ref{BVproblem} to solve a standard boundary value problem in electrostatics, 
to prove that the formulation may have some practical applications. Finally, the merits and the 
demerits of the derived formula will be discussed in section~\ref{discussion}.

\section{Previous Attempts}\label{LuoLiu}
\subsection{Luo Enze's work}
Luo derived a relation between the charge density $\sigma(\bf{r})$ on a conductor and the local 
mean curvature $\kappa_M(\bf {r})$ based on Thomson’s equation Eq.\eqref{1}. He integrated this 
equation along a contour coincident with an electric field line (which is also the direction of 
mean curvature vector of the local equipotential surface), emerging from any point on the conducting surface. 
Of course, the mean curvature of the equipotential surfaces does not remain constant as we move 
a finite distance away from the conductor. However, within an infinitesimal distance from the given conducting surface, it may be 
taken as not to be varying too rapidly. Hence, Luo calculated the electric field very near the 
surface of the conductor~\cite{LuoEnze2}, assuming that the mean curvature of the evolving equipotential surface 
would not vary at all within this infinitesimal distance. He showed that:

\begin{equation}\label{2}
E=E_0\ exp(-2\kappa_M n)
\end{equation}
-where $E_0$ is the field at the conductor surface. This relation becomes more accurate in the 
limit $n\rightarrow0$. $E_0$ is found from the potential difference $\Delta{V}$ over a small 
length $\Delta{n}$ along the field line. Luo determined its value to be:
\begin{equation}\label{3}
E_0=\left(\frac{2\kappa_M\Delta{V}}{exp(-2\kappa_M\Delta{n})-1}\right)_{\Delta{n}\rightarrow0}
\end{equation}
Eq.\eqref{3} was used to obtain the charge density as a function of mean curvature:
\begin{align}\label{4}
\sigma=\epsilon{E_0}&=\left(\frac{2\epsilon\kappa_M\Delta{V}}{exp(-2\kappa_M{\Delta{n}})-1}\right)_{\Delta{n}\rightarrow0}\nonumber\\
                    &\approx\frac{2\epsilon\kappa_M\Delta{V}}{exp(-2\kappa_M{\Delta{n}})-1}
\end{align}
Luo reported the last equation as the desired charge density - curvature relation. Although his 
results are based on few approximations, they satisfy the experimentally observed facts to good 
accuracy~\cite{enze1986distribution}. However, the relation is not very practical as it depends 
on the accurate choice of $\Delta n$ and $\Delta V$. 

Zhang~\cite{zhang1988comment} argued that Luo's calculation was flawed as he had taken constant 
value of mean curvature. The assumption was indeed incorrect. However, the present author finds 
weakness in the illustration which Zhang used to argue the same in his comment letter. If ${\bf 
r}$ is taken to be very near the conducting surface ${\bf r_0}$, then from Eq.(3) of the letter: 

\begin{align}
 E=E_0\frac{r_0^2}{r^2}=E_0\frac{r_0^2}{(r_0+n)^2}&=E_0\left(1+\frac{n}{r_0}\right)^{-2}\nonumber\\
                                                  &\approx E_0\left(1-2\frac{1}{r_0}n\right)\nonumber\\
                                                  &\equiv E_0\left(1-2\kappa_Mn\right)
\end{align}
Zhang argued that Luo's calculation was flawed, as $E_0\frac{r_0^2}{r^2}$ is different from $E_0 
exp(-2\kappa_Mn)$. However, one can easily see that in the limit $n\rightarrow0$, they are not 
different - up to the first order, as $e^x=1+x+..$.

\subsection{Liu and McAllister}
Liu reported (1987) that if the conductor is an ellipsoid or a hyperboloid of revolution of 
two sheets or an elliptic paraboloid of revolution, then $\sigma({\bf{r}})\propto\kappa_G({\bf{r
}})^{0.25}$ holds on each of these bodies~\cite{liu1987relation} [where $\kappa_{G}$ denotes the 
local Gaussian curvature]. The result was confirmed by McAllister in 1990, who discerned that in 
these special cases, Laplace's equation becomes simply separable and the potential becomes a 
function of a single variable~\cite{mcallister1990conductor}.

Liu parametrized Laplace's equation in terms of a function $f$ such that $f(x,y,z)=k$ characterises an 
equipotential surface. One may assume that the potential $\Phi=\Phi(f(x,y,z))$ where $\Phi$ is a solution 
to Laplace's equation $\nabla^2\Phi=0$. With this assumption, one can expand Laplace's equation to obtain~\cite{smythe1950static}:

\begin{equation}\label{5}
\frac{\frac{d}{df}(\frac{d\Phi}{df})}{\frac{d\Phi}{df}}=-\frac{\nabla^{2}f}{(\nabla f)^2} 
\end{equation}
Eq.\eqref{5} serves as the starting point of Liu's work and may be referred to as a key relation. 
He integrated Eq.\eqref{5} and obtained:

\begin{equation}\label{6}
\Phi = A\int e^{-\int \frac{\nabla^2 f}{(\nabla f)^2} df} df + B
\end{equation}
Hence, he calculated particular solution $\Phi(\bf r)$ for each of the conducting bodies that he 
chose to examine. Then, he obtained the electric field by calculating the gradient of $\Phi$ and 
deduced the charge density $\sigma$ at the surface of each of those conducting bodies. Also, the 
Gaussian curvatures of the surfaces of these bodies were calculated by him. Then, he showed that 
for each of these bodies, $\sigma({\bf{r_0}})\propto\kappa_{G}({\bf{r_0}})^{0.25}$. 

McAllister worked with a general orthogonal treatment of Laplace's equation. He showed that when 
the potential is a function of a single variable, it is possible to show that the electric field, 
and therefore $\sigma({\bf{r_0}})$, is proportional to $\frac{1}{\sqrt{g_{ii}}}$, where $g_{ii}$ 
is the square of the scale factor $h_i$ of the $i^{th}$ coordinate of the general orthogonal 
coordinate system. Now, in the case of the surfaces examined by Liu, i.e. ellipsoid, hyperboloid 
of two sheets and elliptic paraboloid, $g_{ii}\propto\frac{1}{\sqrt\kappa_G}$. Thus, the overall 
effect is that the charge density $\sigma({\bf r_0})\propto\kappa_G({\bf r_0})^{0.25}$ for these 
surfaces.

\section{Connection between the existing literature}\label{Bridge}
\subsection{Liu's key relation from Luo's Approach}
The starting points and approaches are apparently very different between the works of Luo and 
Liu, though both of them intend to reach the same goal: finding the dependence of charge density 
on the local surface curvature for an isolated conductor. Luo gives $\sigma$ as a function of 
mean curvature $\kappa_M$ whereas Liu gives $\sigma$ in terms of Gaussian curvature $\kappa_G
$. There should be a connection between these two approaches. To explore that, we parametrize 
Thomson's theorem Eq.\eqref{1} in terms of the (equipotential) surface function $f({\bf{r}})$ 
and expand the L.H.S. as below:

\begin{align*}
 \frac{dE}{ds} + 2\kappa E & = \frac{\partial}{\partial{\bf{r}}}\left(\left|\frac{\partial\Phi}{\partial{\bf{r}}}\right|\right)\cdot\frac{d{\bf{r}}}{ds} + 2\kappa\left|\frac{\partial\Phi}{\partial{\bf{r}}}\right|\\
                           & = {\bf{\hat{n}}}\cdot\frac{\partial}{\partial{\bf{r}}}\left(\left|\frac{\partial\Phi}{\partial f}\frac{\partial f}{\partial{\bf{r}}}\right|\right) + 2\kappa\left(\left|\frac{\partial\Phi}{\partial f}\frac{\partial f}{\partial{\bf{r}}}\right|\right)\\
                           & = {\bf{\hat{n}}}\cdot\left[\left(\frac{\partial}{\partial{\bf{r}}}\frac{\partial\Phi}{\partial f}\right)\left|\frac{\partial f}{\partial{\bf{r}}}\right|+\frac{\partial\Phi}{\partial f}{\frac{\partial}{\partial{\bf{r}}}}\left|\frac{\partial f}{\partial{\bf{r}}}\right|\right] + 2\kappa\left(\frac{\partial\Phi}{\partial f}\left|\frac{\partial f}{\partial{\bf{r}}}\right|\right)\\
                           & = {\bf{\hat{n}}}\cdot\frac{\partial^2\Phi}{\partial f^2}\frac{\partial f}{\partial{\bf{r}}}\left|\frac{\partial f}{\partial{\bf{r}}}\right| + \frac{\partial\Phi}{\partial f}\ {\bf{n}}\cdot\nabla\left|\frac{\partial f}{\partial{\bf{r}}}\right|+\frac{\partial\Phi}{\partial f}\left({\bf\nabla}\cdot{\bf{\hat{n}}}\right)\left|\frac{\partial f}{\partial{\bf{r}}}\right|\\
                           & = \frac{d^2\Phi}{df^2}(\frac{\partial f}{\partial{\bf{r}}}\cdot\frac{\partial f}{\partial{\bf{r}}}) + \frac{d\Phi}{df}({\bf\nabla}\cdot\frac{\partial f}{\partial{\bf{r}}})\\
                           & = \frac{d^2\Phi}{df^2}(\frac{\partial f}{\partial{\bf{r}}})^2 + \frac{d\Phi}{df}(\frac{\partial^2 f}{\partial{\bf{r}}^{2}})\\
                           & = 0
\end{align*}
where we have used $\frac{\partial f}{\partial{\bf{r}}}=|\frac{\partial f}{\partial{\bf{r}}}
|{\bf{n}}$. After some rearrangements, we obtain Eq.\eqref{5}. Liu found it by parametrizing 
Laplace's equation in terms of equipotential surface function $f({\bf r})$. Thus, we reached 
starting point of Liu's work starting from that of Luo's work (i.e. Thomson's theorem). Both 
of them employ Laplace's equation at the fundamental level; because, Eq.\eqref{1} is a consequence 
of Laplace's equation~\cite{matehkolaee2013review}.

\subsection{Luo's formula from Liu's key relation}
Integrating Eq.\eqref{5} between conducting surface equipotential $f_0$ and another arbitrary
equipotential $f(\bf{r})$, we get:

\begin{equation}\label{8}
 \frac{d\Phi}{df}(f({\bf{r}}))=\left(\frac{d\Phi}{df}\right)_{f_0}\ exp\left[-\int_{f_0}^{f(\bf{r})}\frac{\nabla^2f}{|\nabla f|^2}df\right]
\end{equation}
Eq.\eqref{8} may be integrated w.r.t. $f$ to obtain Eq.\eqref{6}. Expanding Eq.\eqref{8} further, we get:

\begin{align}\label{9}
 \frac{d\Phi}{df}(f({\bf{r}})) &= \left(\frac{d\Phi}{df}\right)_{f_0}\ exp\left[-\int_{\bf{r_0}}^{\bf{r}}\frac{\nabla^2f}{|\nabla f|^2} \left(|\nabla f|{\bf{n}}\cdot d{\bf{r}}\right)\right]\nonumber\\
                               &= \left(\frac{d\Phi}{df}\right)_{f_0}\ exp\left[-\int_{\bf{r_0}}^{\bf{r}}\frac{\nabla^2f}{|\nabla f|} {\bf{n}}\cdot d{\bf{r}}\right]
\end{align}
Now, the factor $\frac{\nabla^2f}{|\nabla f|}$ equals:

\begin{align}\label{10}
 \frac{\nabla^2f}{|\nabla f|} &=\frac{1}{|\nabla f|}[\nabla\cdot(|\nabla f|{\bf{n}})]\nonumber\\
                              &=\frac{1}{|\nabla f|}[\nabla|\nabla f|\cdot{\bf{n}}+|\nabla f|\nabla\cdot{\bf{n}}]\nonumber\\
                              &=\frac{\nabla|\nabla f|\cdot{\bf{n}}}{|\nabla f|} + 2\kappa_M\ \ \ \ \ (\rm{as}\ \nabla\cdot{\bf{n}}=2\kappa_M)
\end{align}
Using Eq.\eqref{10}, the integral in the argument of Eq.\eqref{8} becomes:

\begin{align}\label{11}
 \int_{\bf{r_0}}^{\bf{r}}\frac{\nabla^2f}{|\nabla f|}{\bf{n}}\cdot d{\bf{r}} &= \int_{\bf{r_0}}^{\bf{r}}\left[2\kappa_M + {\bf{n}}\cdot\frac{\nabla|\nabla f|}{|\nabla f|}\right]{\bf{n}}\cdot d{\bf{r}}\nonumber\\
							                     &= \int_{\bf{r_0}}^{\bf{r}}2\kappa_M{\bf{n}}\cdot d{\bf{r}}+\int_{\bf{r_0}}^{\bf{r}}{\bf{n}}\cdot\nabla(ln|\nabla f|)\ {\bf{n}}\cdot d{\bf{r}}\nonumber\\
                                                                             &= \int_{\bf{r_0}}^{\bf{r}}2\kappa_M{\bf{n}}\cdot d{\bf{r}}+\int_{\bf{r_0}}^{\bf{r}}\frac{\partial}{\partial{n}}(ln|\nabla f|)\ dn\nonumber\\
                                                                             &= \int_{\bf{r_0}}^{\bf{r}}2\kappa_M{\bf{n}}\cdot d{\bf{r}}+\int_{\bf{r_0}}^{\bf{r}}{d(ln|\nabla f|)}\nonumber\\
                                                                             &= \int_{\bf{r_0}}^{\bf{r}}2\kappa_M{\bf{n}}\cdot d{\bf{r}}+ln\left(\frac{|\nabla f({\bf{r}})|}{|\nabla f({\bf{r_0}})|}\right)
\end{align}
With Eq.\eqref{11}, we can reduce Eq.\eqref{9} to:

\begin{align}\label{12}
 \frac{d\Phi}{df} &= \left(\frac{d\Phi}{df}\right)_{f_0}\ exp\left[-\int_{\bf{r_0}}^{\bf{r}}2\kappa_M{\bf{n}}\cdot d{\bf{r}}-ln\left(\frac{|\nabla f({\bf{r}})|}{|\nabla f({\bf{r_0}})|}\right)\right]\nonumber\\
                  &= \left(\frac{|\nabla f({\bf{r_0}})|}{|\nabla f({\bf{r}})|}\right)\left(\frac{d\Phi}{df}\right)_{f_0}\ exp[-2\int_{\bf{r_0}}^{\bf{r}}\kappa_M{\bf{n}}\cdot d{\bf{r}}]
\end{align}
Now, as $\nabla\Phi({\bf{r}})=\frac{d\Phi(f)}{df}\nabla f({\bf{r}})$, from Eq.\eqref{12}, it follows that:

\begin{equation}\label{13}
 |\nabla\Phi({\bf{r}})|=|\nabla\Phi({\bf{r_0}})|\ exp[-2\int_{\bf{r_0}}^{\bf{r}}\kappa_M{\bf{n}}\cdot d{\bf{r}}]
\end{equation}
Eq.\eqref{13} is the general version of Eq.\eqref{2}. The required charge density is proportional 
to $E_0\equiv|\nabla\Phi({\bf{r_0}})|$ and can be found by integrating Eq.\eqref{13}:

\begin{equation}\label{14}
 |\nabla\Phi({\bf{r_0}})|=\frac{\Phi_0} {\int_\infty^{\bf{r_0}}\ e^{[-2\int_{\bf{r_0}}^{\bf{r}}\kappa_M{\bf{n}}\cdot d{\bf{r}}]}
 {\bf{n}}\cdot d{\bf{r}}}
\end{equation}
In Eq.~\eqref{14}, it is assumed that the reference point where potential $\Phi=0$, is at infinity. 
This need not be the case. In some boundary value problem, if a point ${\bf{r_1}}$ is given, where 
$\Phi=0$, the above equation can then be written as: 
\begin{equation}\label{14a}
 |\nabla\Phi({\bf{r_0}})|=\frac{\Phi_0} {\int_{\bf{r_1}}^{\bf{r_0}}\ e^{[-2\int_{\bf{r_0}}^{\bf{r}}\kappa_M{\bf{n}}\cdot d{\bf{r}}]}
 {\bf{n}}\cdot d{\bf{r}}}
\end{equation}
We shall make use of Eq.~\eqref{14a} in section~\ref{BVproblem}, where the concept will be applied 
to solve a boundary value problem. In rest of the discussions, we shall continue to refer to Eq.~\eqref{14}.

\section{Towards a general relation}\label{universal}
The integral in the denominator of Eq.\eqref{14} cannot be evaluated easily, as the mean curvature 
$\kappa_M$ varies across the continuum of equipotentials and the functional form of this variation 
is unknown. However, the problem can be approached from another way. Let us invoke the Gauss's law 
in electrostatics which asserts that: in a charge free region, the flux of the electrostatic field 
$E\delta A\equiv|\nabla\Phi|\delta A$ is conserved along a flux tube (a bundle of non-intersecting 
field lines perpendicular to the equipotentials). Thus, between two equipotentials at ${\bf{r_0}}$ 
and ${\bf{r}}$:

\begin{equation}\label{15}
 |\nabla\Phi({\bf r})|\delta A({\bf r})=|\nabla\Phi({\bf r_0})|\delta A({\bf r_0})
\end{equation}
Hence, from Eq.\eqref{13} and Eq.\eqref{15}, we have:
\begin{equation}\label{16}
 \delta A({\bf r_0})=\delta A({\bf r})\ exp[-2\int_{\bf{r_0}}^{\bf{r}}\kappa_M{\bf{n}}\cdot d{\bf{r}}]
\end{equation}
Therefore, using Eq.\eqref{16}, Eq.\eqref{14} reduces to:

\begin{equation}\label{17}
 |\nabla\Phi({\bf{r_0}})|=\frac{\Phi_0}{\delta A({\bf{r_o}})\int_{\infty}^{\bf{r_o}}\left[\frac{{\bf{n}}\cdot{d{\bf{r}}}}{\delta A({\bf{r}})}\right]}
\end{equation}
The dependence of the charge density on surface parameters must be obtained by studying Eq.\eqref{17}. 
We notice that the charge density may not be dependent {\it only} on the surface curvature factor(s). 
It can happen that apart from some curvature factors, the charge density (or $|\nabla\Phi({\bf{r_0}})
|$) {\it also} depends on some other functions of the surface coordinates. Therefore, the problem has 
more than one aspects. First, {\it how} is the charge density related to the curvature of the surface 
and second, does this relation completely specify the dependence of the charge density on the surface 
coordinates? In any case, how does it connect to the observation made by Liu and McAllister? 

To answer these questions, the dependence of $\delta A({\bf r_o})$ on the curvature of the conducting 
surface $f({\bf r_0})$ needs to be studied and $\frac{1}{\delta A({\bf{r}})}$ must be integrated over 
a continuum of equipotentials, along an electric field line, starting from a point ${\bf r_0}$ on the 
conducting surface and reaching up to infinity. Evidently, obtaining a formula for the charge density 
effectively means having a method to calculate the electric field when some amount of charge is given 
to a conductor of {\it any} shape. There is no doubt that it is an extremely difficult task.

The area element can be expressed in terms of the coefficients ($E, G, F$) of the first fundamental 
form~\cite[p.~201]{frankel2011geometry} of the conducting surface as: 

\begin{equation}\label{18}
 \delta A({\bf r_0})=\sqrt{EG-F^2}\ \delta u\delta v
\end{equation}
-when the surface is parametrized in terms of $(u,v)$. By invoking the Brioschi formula~\cite[section~
1.5.2]{sternberg2012curvature}, we find:

\begin{align}\label{19}
 \frac{1}{\delta A({\bf r_0})}&\propto\frac{1}{\sqrt{EG-F^2}}\nonumber\\
			      &=\left[{\frac{\kappa_G}
       {
\begin{vmatrix}
-\frac{1}{2}E_{vv}+F_{uv}-\frac{1}{2}G_{uu} & \frac{1}{2}E_u & F_u-\frac{1}{2}E_v\\ 
F_v-\frac{1}{2}G_u & E & F\\
\frac{1}{2}G_v & F & G
\end{vmatrix}
       -
\begin{vmatrix}
0 & \frac{1}{2}E_v & \frac{1}{2}G_u\\ 
\frac{1}{2}E_v & E & F\\
\frac{1}{2}G_u & F & G
\end{vmatrix}
       }
       }
       \right]^{0.25}
\end{align}
where the subscripts $u, v$ denote derivatives with respect to the corresponding parameters. Hence, using 
Eq.\eqref{17} and Eq.\eqref{19}, we find that charge density $\sigma({\bf r_0})$ is proportional to:

\begin{align}\label{20}
\frac{\Phi_0}{\int_\infty^{\bf r_0}\left[\frac{{\bf{n}}\cdot{d{\bf{r}}}}{\delta A({\bf{r}})}\right]}
\left[{\frac{\kappa_G}
       {
\begin{vmatrix}
-\frac{1}{2}E_{vv}+F_{uv}-\frac{1}{2}G_{uu} & \frac{1}{2}E_u & F_u-\frac{1}{2}E_v\\ 
F_v-\frac{1}{2}G_u & E & F\\
\frac{1}{2}G_v & F & G
\end{vmatrix}
       -
\begin{vmatrix}
0 & \frac{1}{2}E_v & \frac{1}{2}G_u\\ 
\frac{1}{2}E_v & E & F\\
\frac{1}{2}G_u & F & G
\end{vmatrix}
       }
       }
       \right]^{0.25}
\end{align}
Clearly, the dependence of $\sigma({\bf r_0})$ on the curvature of the surface comes through the factor $
\kappa_G^{0.25}$ in the numerator of Eq.\eqref{20}. Apart from this dependence, $\sigma({\bf{r_0}})$ also 
depends on (a) the integral factor $\int_\infty^{\bf{r_0}}$ and (b) the difference of the determinants in 
the denominator of Eq.\eqref{19}. All three factors contribute to the charge distribution on a conducting 
surface. The dependence on $\kappa_G^{0.25}$ is generic to all surfaces. The dependence on the difference 
of the determinants in the denominator of Eq.\eqref{19} is also generic to all surfaces. However, for 
different surfaces of different shapes, this factor assumes different values. If the equation 
of the conducting surface $f({\bf{r}}(u,v))$ specified in the problem is known, this factor can be easily 
calculated. The integral factor: $\int_\infty^{\bf{r_0}}$ concerns the equipotentials of the problem. The 
contributions of these surfaces to the integral vary, depending on the presence or absence of other 
charged bodies nearby. Even if an isolated conductor is considered, the calculation of the integral seems 
impossible, because the variation of elementary area $\delta A({\bf r})$ across the equipotentials (from ${\bf r_0}$ to $\infty$) is not 
known. Away from the conductor surface the shape of the equipotential is modified significantly. Although 
the function of the equipotential surface still obeys Eq.\eqref{5}, the variation of $\delta A({\bf r})$ remains unknown. Unless the symmetry of 
the problem allows the equipotentials to be parallel to each other (as in the case of planar, cylindrical 
or spherical symmetries), the exact evaluation of the integral seems impossible. However, even if we keep 
aside the integral factor $\int_\infty^{\bf r_0}$ from discussion for the time being, certain interesting 
aspects of the problem can be addressed from the dependence of the charge density on $\delta A({\bf r_0})
$ which led to Eq.\eqref{19}.

First of all, the dependence of the charge density on $\kappa_{G}^{0.25}$ appeared without the assumption 
that the potential $\Phi({\bf r})$ must be a function of a single variable. However, the density does not 
only depend on the $\kappa_{G}$, but also depends on the function of the surface coordinates, through the 
factor in the denominator of Eq.\eqref{19} and on the integral $\int_\infty^{\bf{r_0}}$. As McAllister illustrated with confocal quadrics and confocal 
paraboloids, the dependency on all these factors reduces to only $\kappa_{G}^{0.25}$, if the potential $\Phi({\bf r})
$ (and therefore, the equipotential surface $f({\bf r})$) becomes a function of a single variable. In all 
the other cases, $\sigma$ continues to depend on all three factors. When $\kappa_{G}$ is 
negative or zero, the denominator also assumes negative or zero value respectively, such that the overall 
factor within the big square bracket (in Eq.\eqref{20}) remains positive and the charge density remains real. A good example 
of this is provided by a toroidal conducting surface. If we assume that the radius from the centre of the 
hole to the centre of the torus tube is $R_2$, and the radius of the tube is $R_1$ (such that $(R_2>R_1)$
), then in the Cartesian coordinates, the equation of a torus azimuthally symmetric about the $z$ axis is 
given by:

\begin{equation}\label{21}
 (R_2-\sqrt{x^2+y^2})^2+z^2=R_1^2
\end{equation}
For calculating the factor in the denominator of Eq.\eqref{19} (call it: DDD-{\bf D}ifference of the {\bf 
D}eterminants in the {\bf D}enominator), any standard parametrization of the surface may be used. In this 
example, we use the following parametrization:

\begin{align}\label{22}
 x&=(R_2+R_1\cos v)\cos u\nonumber\\
 y&=(R_2+R_1\cos v)\sin u\nonumber\\
 z&=R_1\sin v
\end{align}
-where $u,v\in[0,2\pi)$. This particular surface is interesting, because its Gaussian curvature

\begin{equation}\label{23}
 \kappa_G=\frac{\cos v}{R_1(R_2+R_1\cos v)}
\end{equation}
is positive for $0\le v<\frac{\pi}{2}$ and $\frac{3\pi}{2}\le v<2\pi$, zero for $v=\frac{\pi}{2}, 3\frac{
\pi}{2}$ and negative for $\frac{\pi}{2}\le v<\frac{3\pi}{2}$. The factor DDD was evaluated using Wolfram 
Mathematica-10~\cite{mathematica2012wolfram} using $R_2=5$ and $R_1=4$ (in arbitrary units). The plot for 
$\left(\frac{\kappa_G}{DDD}\right)^{0.25}$ is shown in the following figure~\ref{tor} for $v\in[0,2\pi]$:

\begin{figure}[h]
\begin{center}
\includegraphics[width=6.0 cm, height=5.0 cm]{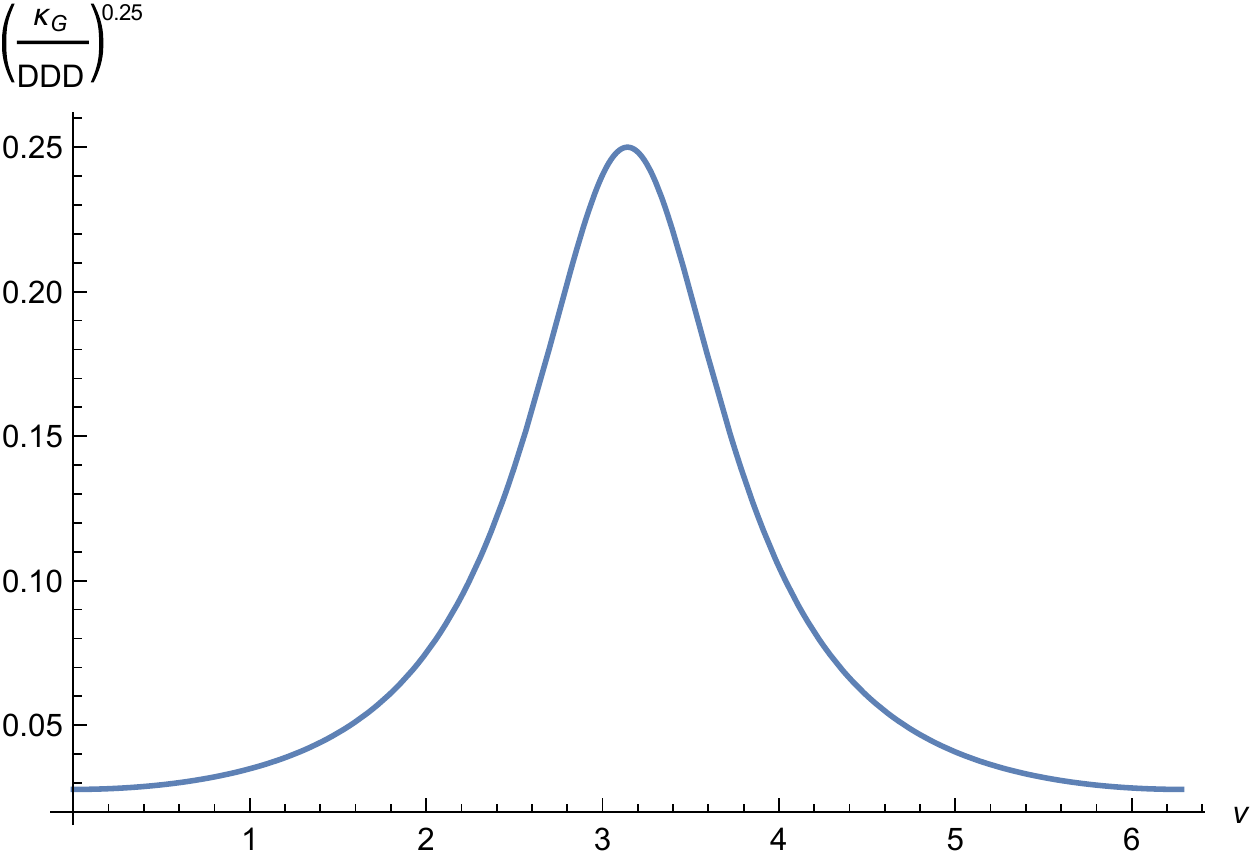}
\caption{The plot for $\left(\frac{\kappa_G}{DDD}\right)^{0.25}$ for torus given by Eq.\eqref{22}}
\label{tor}
\end{center}
\end{figure}
The above plot shows that $\left(\frac{\kappa_G}{DDD}\right)^{0.25}$ always remains positive irrespective 
of the sign or value of $\kappa_G$ of a surface. The charge density on the torus~\cite{torres1989comment}
is a more complicated function of $v$ and is different from the variation shown in figure~\ref{tor}. Thus, 
the {\bf C}ompensating {\bf F}actor (CF) that must be multiplied to $\left(\frac{\kappa_G}{DDD}\right)^{0
.25}$ to obtain the charge density, arises due to the integral factor $\int_\infty^{\bf{r_0}}$. The value 
of this factor can be back-calculated if the charge density is already known, using $\left(\frac{\kappa_G
}{DDD}\right)^{0.25}$. In the following table~\ref{Tab1}, the summary of all the relevant factors is presented 
for a few conducting surfaces.

\begin{table}[ht]
\caption{Analysis of charge density dependence on various factors}\label{Tab1}
\vspace{0.25cm}
\centering
\begin{tabular}{|c|c|c|c|c|}
\hline
{\bf{Surface}}&{\bf{Parametrization}}&{\bf{DDD}}&{\makecell{\bf{$\left(\frac{\kappa}{DDD}\right)^{0.25}$}}}&{\bf{CF}}\\
\hline                             
\makecell{ellipsoid\\$\frac{x^2}{a^2}+\frac{y^2}{b^2}+\frac{z^2}{c^2}=1$}&\makecell{$x=a\cos u\sin v$\\$y=b\sin u\sin v$\\$z=c\cos v$}&$\propto\frac{1}{\sin^4 v}$&$\frac{\kappa^{0.25}}{\sin v}$&$\sin v$\\
\hline
\makecell{hyperboloid of 2 sheets\\$\frac{x^2}{a^2}+\frac{y^2}{b^2}-\frac{z^2}{c^2}=-1$}&\makecell{$x=a\sinh u\cos v$\\$y=b\sinh u\sin v$\\$z=c\cosh u$}&$\propto\frac{1}{\sinh^4 u}$&$\frac{\kappa^{0.25}}{\sinh u}$&$\sinh u$\\
\hline
\makecell{elliptic paraboloid\\$\frac{x^2}{a^2}+\frac{y^2}{b^2}=z$}&\makecell{$x=a\sqrt{u}\cos v$\\$y=b\sqrt{u}\sin v$\\$z=u$}& constant & $\kappa^{0.25}$ & none \\
\hline
\end{tabular}
\end{table}

\section{Application to a boundary value problem}\label{BVproblem}
After the discussion in the preceding section and from Eq.~\eqref{14a}, it is evident that if in a given 
boundary value problem, the point ${\bf r_1}$ where potential $\Phi=0$ is given, then the charge density 
will be proportional to:

\begin{equation}\label{24}
 \frac{\Phi_0}{\delta A({\bf{r_o}})\int_{\bf{r_1}}^{\bf{r_o}}\left[\frac{{\bf{n}}\cdot{d{\bf{r}}}}{\delta A({\bf{r}})}\right]}
\end{equation}
Eq.~\eqref{24} is the same as Eq.~\eqref{17} except for the lower limit of the integral of $\left[\frac{
{\bf{n}}\cdot{d{\bf{r}}}}{\delta A({\bf{r}})}\right]$. Next, we shall apply Eq.~\eqref{24} to a boundary 
value problem and validate the power of this formulation.
\begin{figure}[h]
\begin{center}
\includegraphics[width=6.0 cm, height=5.0 cm]{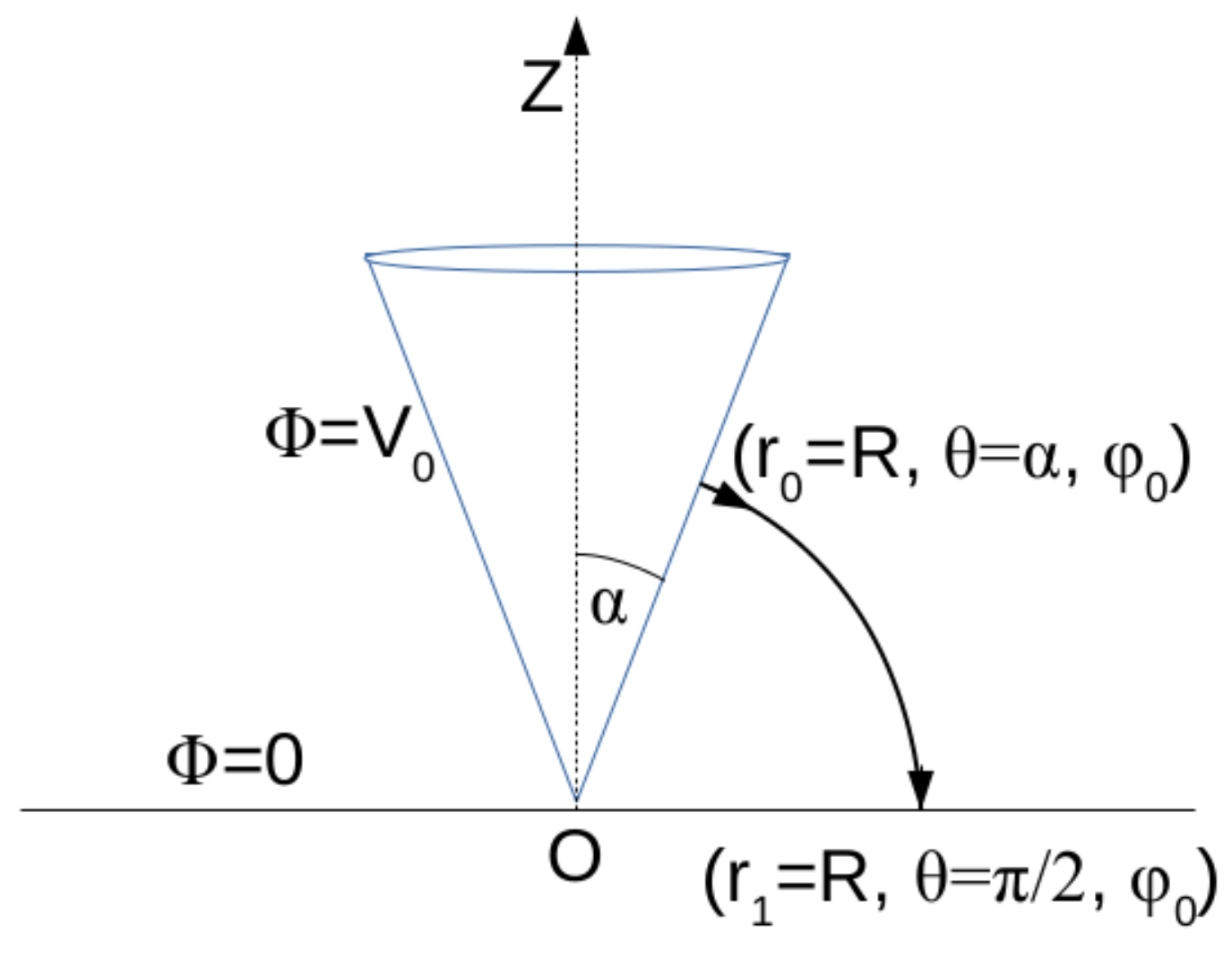}
\caption{Conducting cone just above a grounded conducting plane}
\label{cone}
\end{center}
\end{figure}

We take a classic example of an infinite conducting cone at potential $V_0$ with its apex facing down on 
a grounded conducting plane~\cite{sadiku2007elements}. The cone has half-angle $\alpha$ and is open upwards (see figure~\ref{cone}). 
The apex of the cone is separated from the conducting plane by a tiny air gap (to provide insulation). We 
try to find the functional form of the charge density on the cone.

The standard method of solving this boundary value problem is to solve the zenith angle dependent part of 
the Laplace's equation. This gives the electrostatic potential as a function of $\theta$, i.e. $\Phi=\Phi
(\theta)$. The electric field is the negative gradient of this potential. The value of the field close to 
the given conical conducting surface is proportional to the desired charge density.

However, our formulation directly gives the functional dependence of the electric field, near the conical 
surface. In this case, the equipotentials are all conical surfaces within the region $\alpha<\theta<\frac
{\pi}{2}$. As $\kappa_G=0$ on the conical surface, we calculate $\delta A({\bf r_0})$ from Eq.~\eqref{18}. 
The conical surface is parametrized as:

\begin{align}\label{25}
 x&=\frac{r}{a}\cos\phi\nonumber\\
 y&=\frac{r}{a}\sin\phi\nonumber\\
 z&=r
\end{align}
Then, we can easily show that $\sqrt{EG-F^2}\propto r$. So, at a point ${\bf r_0}[=(R,\alpha,\phi_0)]$ on 
the surface of the cone, we have:
\begin{align}\label{26}
 \delta A({\bf r_0})&\propto R\ dr\ R\sin\alpha d\phi\nonumber\\
		    &=R^2\sin\alpha dr d\phi
\end{align}
If we follow a field line that originates from a patch around ${\bf r_0}$ on the given conducting conical 
surface, we see that it reaches the grounded conducting plane at a point ${\bf r_1}[=(R,\theta=\frac{\pi}
{2},\phi_0)]$. We notice that only a single variable ($\theta$) varies in this problem. Now, the integral 
can be evaluated as:

\begin{align}\label{27}
 \int_{\bf{r_1}}^{\bf{r_o}}\left[\frac{{\bf{n}}\cdot{d{\bf{r}}}}{\delta A({\bf{r}})}\right]&=\int_{\bf{r_1}}^{\bf{r_o}}\left[\frac{R\ d\theta}{R^2\sin\theta\ dr\ d\phi}\right]\nonumber\\
 &=\frac{1}{R\ dr\ d\phi}\int_{\theta=\frac{\pi}{2}}^{\theta=\alpha}\frac{d\theta}{\sin\theta}\nonumber\\
 &=\frac{1}{R\ dr\ d\phi}\tan\frac{\alpha}{2}
\end{align}
Therefore, from Eq.~\eqref{24}, Eq.~\eqref{26} and Eq.~\eqref{27}, the charge density on the conical surface is proportional to:

\begin{equation}\label{28}
 \sigma({\bf r_0})\propto\frac{V_0}{R\sin\alpha\ \rm{ln}(\tan\frac{\alpha}{2})}
\end{equation}
Similarly, one can also derive the charge density on the grounded conducting plane and its value is found 
to be proportional to $\frac{V_0}{R\ \rm{ln}(\tan\frac{\alpha}{2})}$ at a distance $R$ from the origin $O
$. It is also possible to directly calculate the electric field in the region $\alpha<\theta<\frac{\pi}{2
}$ using Eq.~\eqref{15}. As stated earlier, this formulation is more direct in calculating charge density, 
compared to the conventional methods. 

\section{Discussions}\label{discussion}
In this paper, we reviewed the literature on charge density dependence on conductor surface curvature 
and tried to establish that these works, though very different in their approaches, are actually related 
to each other. A number of authors indicated the weak points in these works but the literature were 
overall silent about how to generalise these results. The connection between the work of Luo~\cite{enze1986distribution} 
and Liu~\cite{liu1987relation} was also not well understood. A recent pedagogical article~\cite{matehkolaee2013review} 
also could not throw light on this topic. 

In this paper, we clarified that the charge density partially depends on $0.25^{th}$ power of the Gaussian 
curvature of the given conducting surface (through Eq.\eqref{20}), though the mean curvature $\kappa_M$ of 
the equipotentials appears in the integral (from ${\bf{r_0}}$ to ${\bf{r}}$) in the expression of electric 
field near the conducting surface in Eq.\eqref{14}. It is an important point, because the $\kappa_G$ is an 
intrinsic property of a surface whereas $\kappa_M$ is not. However, our work indicates that charge density 
{\it also} depends on other functions dependent on surface coordinates. Thus, the dependence on $\kappa_G$
cannot be used always to know the distribution of charge on a given conducting surface. In some cases like 
elliptic paraboloid, it may happen that the contributions from DDD and CF are constants and $\sigma\propto
\kappa_G^{0.25}$ - but that does not happen in general.

We have shown that when the reference point of the potential is at a finite distance, the integral $\int_{
\bf{r_1}}^{\bf{r_0}}\frac{{\bf n\cdot}d{\bf r}}{\delta A({\bf r})}$ can be evaluated if certain symmetries 
permit. In this case, the formulation offers a direct way to find the functional dependence of the density 
of charge. However, if the symmetries do not permit, evaluation of the integral $\int_{\bf r_1}^{\bf r
_0}$ (or $\int_\infty^{\bf{r_0}}$) seems to be an impossible task, even when the potential depends on a 
single variable, because the area element $\delta A({\bf r})$, ${\bf n}({\bf r})$ and $d{\bf r}$ gradually 
evolve along with the different equipotentials. The contribution of this factor into the charge density is 
quantified by the compensating factor CF. For the examples in table~\ref{Tab1}, it appears that CF somehow 
depends on $\frac{dz}{d\xi}$, where $z$ is a function of $\xi$ (in the parametric equation of the surface). 
But this empirical observation should not be interpreted as a general rule.



\section{Acknowledgements}
I feel indebted to Dr. Debapriyo Syam for precious encouragements from him while trying to solve the 
problem. The supports from from my friends Tanmay, Manoneeta and Tamali were also of great help.

\bibliographystyle{unsrt}
\bibliography{ES_charge}

\begin{thebibliography}{10}

\bibitem{griffiths1999introduction}
David~Jeffrey Griffiths and Reed College.
\newblock {\em Introduction to electrodynamics}, volume~3.
\newblock prentice Hall Upper Saddle River, NJ, 1999.

\bibitem{greiner2012classical}
Walter Greiner.
\newblock {\em Classical electrodynamics}.
\newblock Springer Science \& Business Media, 2012.

\bibitem{green1889essay}
George Green.
\newblock {\em An essay on the application of mathematical analysis to the
  theories of electricity and magnetism}, volume~3.
\newblock author, 1889.

\bibitem{bakhoum2008proof}
Ezzat~G Bakhoum.
\newblock Proof of thomson's theorem of electrostatics.
\newblock {\em Journal of Electrostatics}, 66(11):561--563, 2008.

\bibitem{estevez1985power}
GA~Estevez and LB~Bhuiyan.
\newblock Power series expansion solution to a classical problem in
  electrostatics.
\newblock {\em American Journal of Physics}, 53(2):133--134, 1985.

\bibitem{pappas1986differential}
Richard~C Pappas.
\newblock Differential-geometric solution of a problem in electrostatics.
\newblock {\em SIAM Review}, 28(2):225--227, 1986.

\bibitem{enze1986distribution}
Luo Enze.
\newblock The distribution function of surface charge density with respect to
  surface curvature.
\newblock {\em Journal of Physics D: Applied Physics}, 19(1):1--6, 1986.

\bibitem{liu1987relation}
Kun-Mu Liu.
\newblock Relation between charge density and curvature of surface of charged
  conductor.
\newblock {\em American Journal of Physics}, 55(9):849--852, 1987.

\bibitem{mcallister1990conductor}
IW~McAllister.
\newblock Conductor curvature and surface charge density.
\newblock {\em Journal of physics. D, Applied physics}, 23(3):359--362, 1990.

\bibitem{dube1989comment}
Myriam Dub{\'e}, Mario Morel, N~Gauthier, and AJ~Barrett.
\newblock Comment on``relation between charge density and curvature of surface
  of charged conductor,''by kun-mu liu [am. j. phys. 55, 849-852 (1987)].
\newblock {\em American Journal of Physics}, 57:1047--1048, 1989.

\bibitem{torres1989comment}
M~Torres, JM~Gonz{\'a}lez, and G~Pastor.
\newblock Comment on``relation between charge density and curvature of surface
  of charged conductor,''by kun-mu liu [am. j. phys. 55, 849-852 (1987)].
\newblock {\em American Journal of Physics}, 57:1044--1046, 1989.

\bibitem{zhang1988comment}
Yuan~Zhong Zhang.
\newblock A comment on'the distribution function of surface charge density with
  respect to surface curvature'.
\newblock {\em Journal of Physics D: Applied Physics}, 21(7):1235, 1988.

\bibitem{LuoEnze2}
Luo Enze.
\newblock The application of a surface charge density distribution function to
  the solution of boundary value problems.
\newblock {\em Journal of Physics D: Applied Physics}, 20(1):1609--1615, 1987.

\bibitem{smythe1950static}
William~Ralph Smythe and William~R Smythe.
\newblock {\em Static and dynamic electricity}, volume~3.
\newblock McGraw-Hill New York, 1950.

\bibitem{matehkolaee2013review}
Mehdi~Jafari Matehkolaee and Ali~Naderi Asrami.
\newblock The review on the charge distribution on the conductor surface.
\newblock {\em European J Of Physics Education}, 4(3), 2013.

\bibitem{frankel2011geometry}
Theodore Frankel.
\newblock {\em The geometry of physics: an introduction}.
\newblock Cambridge University Press, 2011.

\bibitem{sternberg2012curvature}
Shlomo Sternberg.
\newblock {\em Curvature in mathematics and physics}.
\newblock Courier Corporation, 2012.

\bibitem{mathematica2012wolfram}
Wolfram Mathematica.
\newblock Wolfram research.
\newblock {\em Inc., Champaign, Illinois}, 2012.

\bibitem{sadiku2007elements}
Matthew~NO Sadiku.
\newblock {\em Elements of electromagnetics}.
\newblock Oxford university press, 2007.

\end{thebibliography}
\end{document}